\documentclass{PoS}

\def\al{\alpha}
\def\be{\beta}

\def\de{\delta}

\def\la{\lambda}

\def\si{\sigma}

\def\ta{\tau}

\def\ch{\chi}
\def\ps{\psi}
\def\om{\omega}
\def\Ga{\Gamma}
\def\De{\Delta}

\def\nue{\nu_e}
\def\numu{\nu_\mu}

\def\nuebar{\bar\nu_e}
\def\numubar{\bar\nu_\mu}

\def\fr#1#2{\frac{#1}{#2}}

\def\lsim{\mathrel{\rlap{\lower4pt\hbox{\hskip1pt$\sim$}}
    \raise1pt\hbox{$<$}}}
\def\gsim{\mathrel{\rlap{\lower4pt\hbox{\hskip1pt$\sim$}}
    \raise1pt\hbox{$>$}}}

\newcommand{\beq}{\begin{eqnarray}}
\newcommand{\eeq}{\end{eqnarray}}
\def\to{\rightarrow}

\def\no{\nonumber}

\def\As#1{({\cal A}_s)_{#1}}
\def\Ac#1{({\cal A}_c)_{#1}}
\def\Bs#1{({\cal B}_s)_{#1}}
\def\Bc#1{({\cal B}_c)_{#1}}
\def\C#1{({\cal C})_{#1}}

\def\nh^#1{{\hat N}^{#1}}

\def\mF_#1{{\cal F}_{#1}} 

\def\gaf{{\gamma}_5}
\def\gam{{\gamma}_{\mu}}

\def\ganu{{\gamma}^{\nu}}

\def\MBosc1POT{5.58\times 10^{20}}

\def\tsid{86164.1}
\def\tgmt{86400.0}

\def\Dcsqnu{26.9}
\def\Dcsqnubar{3.0}

\title{Tests of Lorentz and CPT violation with neutrinos}

\ShortTitle{Tests of Lorentz and CPT violation with neutrinos}

\author{Teppei Katori
        \thanks{On behalf of LSND, MiniBooNE, 
        and Double Chooz collaborations}\\
        Laboratory for Nuclear Science,\\ 
        Massachusetts Institute of Technology,\\
        Cambridge, USA\\
        E-mail: \email{katori@mit.edu}}

\abstract{
Lorentz violation is a predicted phenomenon from the Planck scale physics. 
Although the three active massive neutrino framework with the Standard Model (SM), 
so-called the neutrino Standard Model ($\nu$SM), is successful, 
series of signals not understood within the $\nu$SM suggest neutrino physics 
may be the first place to see the physics beyond the SM, 
such as Lorentz violation. 
Especially, neutrino oscillations are the natural interferometer and 
they are sensitive to the Lorentz violation with comparable sensitivity 
with precise optical experiments. 

The LSND oscillation signal was analyzed under 
the Standard Model Extension (SME) framework, 
and it was found that the oscillation data was consistent with no Lorentz violation, 
but data cannot reject Lorentz violation hypothesis with order $\sim10^{-17}$. 
By assuming LSND signal was due to the Lorentz violation, 
a global phenomenological model was made to describe all known oscillation 
data including the LSND signal. 
The model also predicted the signal for MiniBooNE at the low energy region. 

Later, MiniBooNE announced an event excess at the low energy region. 
However, the oscillation candidate signals from MiniBooNE were 
consistent with no Lorentz violation. 
The limit obtained by MiniBooNE and MINOS on the $e-\mu$ sector reject 
the simple scenario to explain LSND signal with Lorentz violation.

Meantime, MINOS and IceCube set tight limits on the $\mu-\tau$ sector Lorentz violation. 
The last untested channel, the $e-\tau$ Lorentz violating mixing, 
was tested using reactor disappearance data from Double Chooz. 
However, Double Chooz data was consistent with flat, 
and sidereal time dependent Lorentz violation hypothesis is rejected. 
Combinations of all oscillation data from LSND, MiniBooNE, MINOS, IceCube, 
and Double Chooz provide very tight constraint for 
a possible Lorentz violation in the neutrino sector in terrestrial level.
}

\FullConference{36th International Conference on High Energy Physics,\\
		July 4-11, 2012\\
		Melbourne, Australia}

\begin{document}

\section{Introduction}

\subsection{Spontaneous Lorentz Symmetry Breaking (SLSB)}

\begin{figure}[tbp]
\begin{center}
\includegraphics[width=10cm]{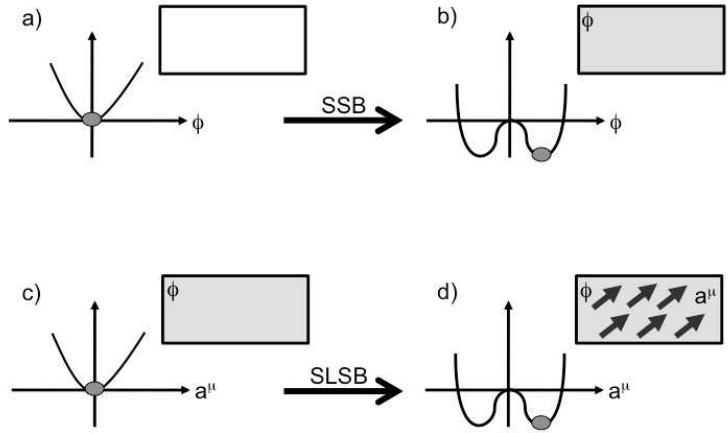}
\end{center}
\caption{\label{fig:SLSB}
An illustration of spontaneous symmetry breaking (SSB). 
Figure is taken from ~\cite{lv_review}.
}
\end{figure}

Lorentz violation is a predicted phenomenon from the Planck scale physics. 
Especially if it were made by the spontaneous process, 
quantum field theory and general relativity would require no modifications. 
Figure~\ref{fig:SLSB} shows a cartoon of this situation. 
When the universe is hot, 
the scalar field preserves the perfect symmetry (Fig.~\ref{fig:SLSB}a). 
But once it gets cold, 
there is a chance that the potential of 
the scalar field shifts to the "Mexican hat" potential (Fig.~\ref{fig:SLSB}b), 
and nonzero field value is more stable, 
{\it i.e.} the vacuum acquires the vacuum expectation value of this scalar field. 
If the scalar field has any quantum numbers, 
say SU(2) charge, such quantum numbers are not preserved in the vacuum, 
namely chirality is not conserved for massive particles in the SM vacuum. 
This is the spontaneous symmetry breaking (SSB) in the SM, 
and this year is the great year for this mechanism since 
the strong candidate of Higgs particle is discovered, 
and presented at this ICHEP2012 conference! 

This process can be extended to any field with Lorentz indices beyond the scalar field. 
For simplicity I discuss only the vector field. 
When the universe is hot, vector field keeps the perfect symmetry (Fig.~\ref{fig:SLSB}c). 
But again, when universe gets cold, 
the vector field could generate nonzero vacuum expectation values (Fig~\ref{fig:SLSB}d). 
This is the situation of the spontaneous Lorentz symmetry violation (SLSB)~\cite{SLSB}, 
and Lorentz symmetry is spontaneously broken. 
In this case, the universe is filled with 
the background vector fields represented by the arrows. 
If the SM particles couple with those fields, 
their physics depends on the orientation of the arrows or direction. 

Since such background fields of the universe are fixed in the space, 
presence of such coupling implies the direction dependent physics. 
In particular, rotation of the Earth (period $\tsid$~sec) 
causes sidereal time dependent physics for terrestrial measurements. 
Therefore, sidereal time dependence of physics observables 
is the smoking gun of the Lorentz violation. 

\subsection{Particle and Observer Lorentz transformation}

\begin{figure}[tbp]
\begin{center}
\includegraphics[width=10cm]{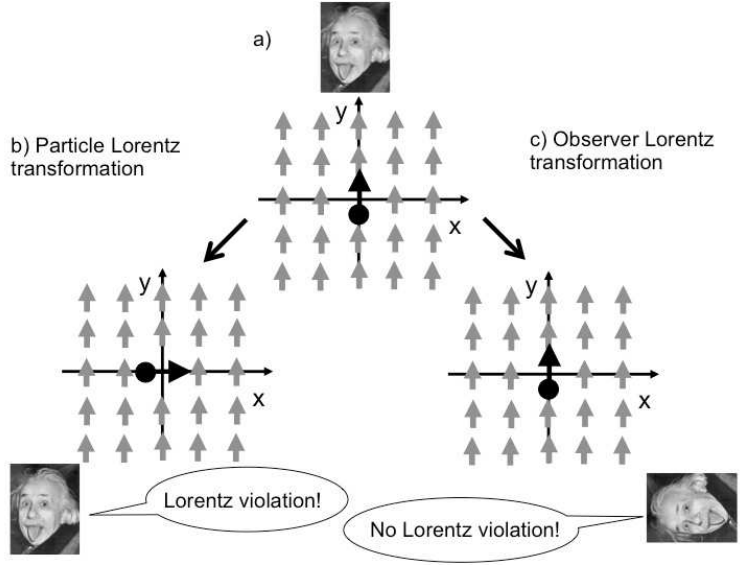}
\end{center}
\caption{\label{fig:LT}
An illustration of the Particle Lorentz transformation and 
the Observer Lorentz transformation.
 Figure is taken from~\cite{lv_review}.
}
\end{figure}

Lorentz violation is more precisely the violation of the Particle Lorentz transformation. 
This situation is described in Figure~\ref{fig:LT}. 
In Fig. ~\ref{fig:LT}a, 
a SM particle is moving upward in two-dimensional space where 
the hypothetical background vector field saturates the space (depicted by arrows), 
and Einstein represents a local observer. 

The change in motion of a SM particle within a fixed coordinate system is described by 
Particle Lorentz transformation (Fig.~\ref{fig:LT}b). 
Since the background field is unchanged, as a consequence, 
a coupling between the SM particle and vector field is not preserved. 
In general, Lorentz violation means Particle Lorentz violation. 

On the other hand, local observer's inverse coordinate change can also 
generate change in the motion of a SM particle (Fig.~\ref{fig:LT}c). 
In the theory without Lorentz violation, 
this Observer Lorentz transformation coincides with the Particle Lorentz transformation. 
However, as you see, this corresponds to mere coordinate transformation and 
the coupling of the SM particle and the background field is preserved. 
Therefore, Lorentz-violating effect is conserved by the coordinate transformation, 
and it can be studied in any frame or coordinate.

\subsection{Test of Lorentz violation with neutrino oscillations}

\begin{figure}[tbp]
\begin{center}
\includegraphics[width=10cm]{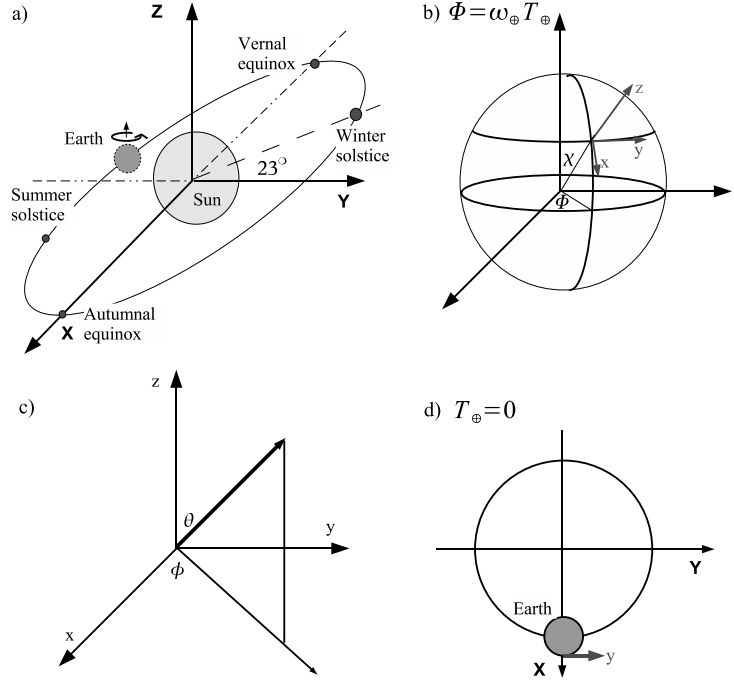}
\end{center}
\caption{\label{fig:coords}
The coordinate system used by this analysis, 
the Sun-centered coordinates (a), 
the Earth-centered coordinates (b), 
and the local polar coordinate system (c). 
The time zero is defined when the experiment site is at 
midnight near the autumnal equinox, 
in other words, when the large ``Y'' and small ``y'' axes almost align (d). 
Figure is taken from~\cite{LSND_LV}.
}
\end{figure}

Lorentz violation is realized as a coupling of SM particles and 
the background field of the universe. 
Although the Lorentz-violating phenomenon is coordinate independent, 
we need to choose a coordinate system so that 
we can report measurements of the coefficients of such fields~\cite{SMEtable}. 
Figure~\ref{fig:coords} shows our scheme. 
First, the motion of the Earth is described in 
the Sun-centered coordinate system (Fig.~\ref{fig:coords}a). 
This coordinate system provides the bases for 
the Lorentz violating fields to specify the coefficients. 
Then the location of the experimental site is specified in 
the Earth-centered coordinate system (Fig.~\ref{fig:coords}b). 
Finally, the direction of SM particles 
({\it i.e.}, direction of the beam) 
is described in the local polar coordinate system (Fig.~\ref{fig:coords}c).

The Standard Model Extension (SME)~\cite{SME} is constructed as 
a general framework to analyze the data to search possible Lorentz violation. 
In the SME, Lorentz violating interactions are described by 
the perturbative terms in the Lagrangian, on top of the SM terms. 
We are particularly interested in 
the test of Lorentz violation using neutrino oscillations. 
Since neutrino oscillation is a natural interferometer, 
small couplings of neutrinos with Lorentz violating fields could cause phase shifts and 
could result in neutrino oscillations. 
The sensitivity of neutrino oscillations to Lorentz violation is 
comparable with precise optical experiments. 
To analyze the neutrino data we use the neutrino sector SME~\cite{KM1}~\footnote{
In this article, we limit outselves within the renormalizable SME only. 
However, the results in this article can also set the limits on 
the nonrenormalizable SME coefficients~\cite{KM4}.
}, 

\beq
{\cal L} 
&=& 
\fr{1}{2}i{\bar {\ps}}_A{\Ga}^{\mu}_{AB}\stackrel{\leftrightarrow}
{D_{\mu}}{\ps}_{B}-{\bar {\ps}}_{A} M_{AB}{\ps}_{B}+h.c.,\\
\Ga_{AB}^{\nu}&\equiv&
\ganu\de_{AB}+c_{AB}^{\mu\nu}\gam+d_{AB}^{\mu\nu}\gaf\gam+e_{AB}^{\nu}+if_{AB}^{\nu}\gaf
+\fr{1}{2}g_{AB}^{\la\mu\nu}\si_{\la\mu},~\label{eq:gamma}\\
M_{AB}&\equiv&
m_{AB}+im_{5AB}\gaf+a_{AB}^{\mu}\gam+b_{AB}^{\mu}
+\fr{1}{2}H_{AB}^{\mu\nu}\si_{\mu\nu}.~\label{eq:mass}
\eeq

Here, the $AB$ subscripts represent flavor space. 
The first term of Eq.~\ref{eq:gamma} and the first and the second terms of
Eq.~\ref{eq:mass} are the only nonzero terms in the SM.  
The rest of the terms are from the SME. 
These SME coefficients can be 
classified into two groups: 
$e^{\mu}_{AB}$, $f^{\mu}_{AB}$,
$g^{\mu \nu \la}_{AB}$, $a^{\mu}_{AB}$, and $b^{\mu}_{AB}$ 
are CPT-odd SME coefficients, 
and $c^{\mu \nu}_{AB}$, $d^{\mu \nu}_{AB}$, and $H^{\mu\nu}_{AB}$ 
are CPT-even SME coefficients. 

In this way, physical observables can be written down including Lorentz violation. 
By assuming the baseline is short enough for the oscillation length
the neutrino oscillation probability $\al\to\be$ 
can be written as follows~\cite{KM3}~\footnote{
If this is not the case, Lorentz violation can be studied as 
perturbations of standard oscillations~\cite{LBA}. 
Undermentioned MINOS far detector and IceCube analyses are based on this scheme. 
}, 

\beq
P_{\al\to\be} & \simeq & \fr{L^2}{(\hbar c)^2} |\, \C{\al\be} 
  +\As{\al\be} \sin\om_\oplus T_\oplus
  +\Ac{\al\be} \cos\om_\oplus T_\oplus \no\\
 & & 
  +\Bs{\al\be} \sin2\om_\oplus T_\oplus
  +\Bc{\al\be} \cos2\om_\oplus T_\oplus\,|^2.
\label{eq:nu_prob}
\eeq

Here, $\om_\oplus$ is the sidereal time angular frequency 
($\om_\oplus=\frac{2\pi}{\tsid}$~rad/s).  
The neutrino oscillation probability is described by 
the function of the sidereal time $T_\oplus$ with five amplitudes. 
$\C{\al\be}$ is the sidereal time independent amplitude, and 
$\As{\al\be}$, $\Ac{\al\be}$, $\Bs{\al\be}$, and $\Bc{\al\be}$ 
are the sidereal time dependent amplitudes. 
Therefore, an analysis of Lorentz and CPT violation in neutrino oscillation data involves 
fitting the data with Eq.~\ref{eq:nu_prob} to 
find nonzero sidereal time dependent amplitudes. 
These amplitudes are written by a combination of the SME coefficients, 
and the explicit expressions are given at elsewhere~\cite{KM3}.

\section{Lorentz violation analysis on LSND experiment}

LSND is an appearance neutrino oscillation experiment at Los Alamos. 
A low energy $\numubar$ beam ($\sim$40~MeV) was made by pion decay-at-rest. 
The detector was located $\sim$30~m away from the target. 
LSND observed the excess of $\nuebar$ candidate events from 
the $\numubar$ beam~\cite{LSND_osc}, 
and this result is not understood within the $\nu$SM. 
Thus, LSND signal may be the signal of new physics, 
such as sterile neutrino oscillations. 
However, it may be the first signal of the Lorentz violation. 
If this is the case, interference pattern, 
{\it i.e.}, the number of the oscillation candidates, 
would depend on the sidereal time. 

\begin{figure}[tbp]
\begin{center}
\includegraphics[width=12cm]{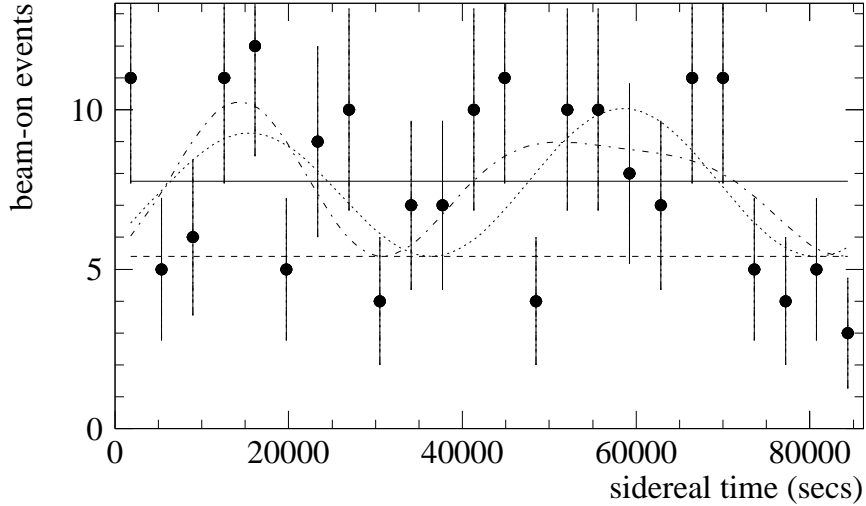}
\end{center}
\caption{\label{fig:LSND}
Sidereal time distribution of the LSND oscillation candidate events (marker). 
Data is fit with oscillation models including sidereal time dependence 
(dotted and dot-dashed curves) 
and flat hypothesis (solid line). 
The background is assumed to be flat (dashed line). 
Figure is taken from~\cite{LSND_LV}.
}
\end{figure}

We analyzed the LSND oscillation candidate data with 
a function of the sidereal time~\cite{LSND_LV}. 
We fit Eq.~\ref{eq:nu_prob} to find the best parameter set by 
an unbinned likelihood method. 
Fig.~\ref{fig:LSND} shows the result. 
The data is consistent with a flat hypothesis (=no sidereal time dependence), 
however small Lorentz violation (=sidereal time dependent solution) is not rejected. 
If that is the case, LSND oscillation candidate is 
explained by order $10^{-17}$ CPT-odd Lorentz violation and/or 
order $10^{-19}$~GeV CPT-even Lorentz violation~\cite{CPT10_TK}.

\section{Global neutrino oscillation model with Lorentz violation}

\begin{figure}[tbp]
\begin{center}
\includegraphics[width=10cm]{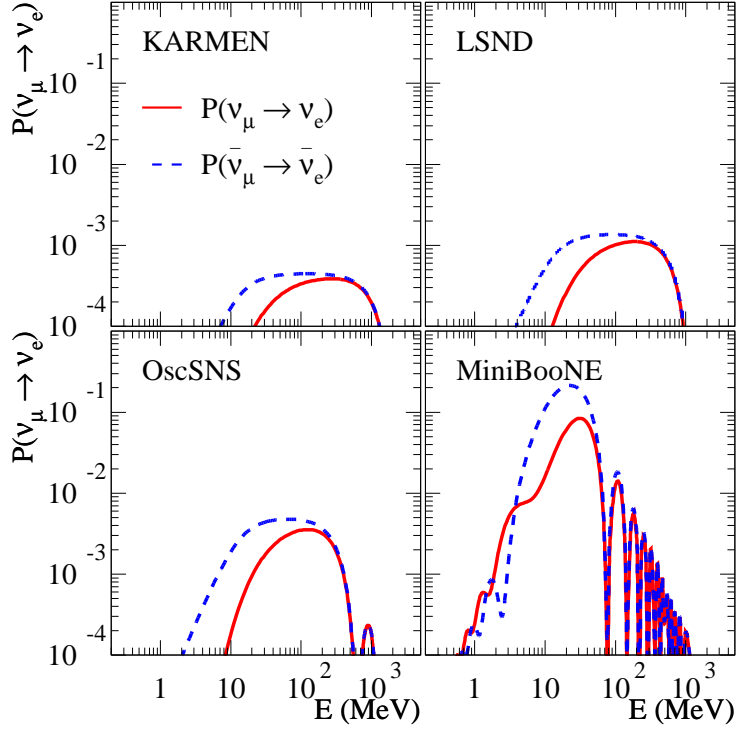}
\end{center}
\caption{\label{fig:tandem}
Oscillation probabilities as a function of energy for neutrino (solid line) 
and antineutrino (dashed lines). 
Figure is taken from~\cite{tandem}.
}
\end{figure}

Although LSND signal could be described by the Lorentz violation, 
naively such new parameters would be forbidden by other oscillation experiment data. 
We examined the possible phenomenological models to 
describe the world oscillation data including LSND. 
The "tandem" model was made in such concept~\cite{tandem} 
(as an extension of the "bicycle" model~\cite{KM2}). 
The tandem model satisfies all requirements as an alternative oscillation model. 
One of the very attractive features of this model is that 
it only uses three parameters to describe four oscillation signals; 
solar, atmospheric, KamLAND, and LSND signals. 
In 2006, the $\nu$SM had four parameters 
(two mass differences and two mixing angles), 
so the tandem model was a more economical phenomenological model than the $\nu$SM. 
Fig.~\ref{fig:tandem} shows our prediction on the short baseline experiments. 
It reproduces a $\sim$0.1\% level oscillation signal at LSND. 
On the other hand, the signal at KARMEN is smaller to be consistent with the observation. 
It also predicted an oscillation signal at the low energy region of MiniBooNE, 
both neutrinos and antineutrinos. 

Later more advanced global oscillation models had been proposed~\cite{puma}, 
but all of them failed to reproduce recent reactor neutrino results~\cite{reactor}, 
which is another great discovery of this year, 
and it is also being presented at this ICHEP2012 conference!

\section{Lorentz violation analysis on MiniBooNE experiment}

\subsection{MiniBooNE experiment}

MiniBooNE is the neutrino and antineutrino appearance experiment 
designed to confirm or reject LSND signals under 
the two massive neutrino oscillation hypothesis. 
The neutrino (antineutrino) beams are created by 
the Booster Neutrino Beamline (BNB)~\cite{MB_beam}. 
The 8~GeV proton beam is extracted from the Fermilab Booster, 
and it is sent to the target where the collision with 
the target makes shower of mesons. 
The magnetic focusing horn surrounding the target selects 
either positive mesons or negative mesons and their decay-in-flight 
make $\sim$800~MeV $\numu$ or $\sim$600~MeV $\numubar$ beam. 

The MiniBooNE detector is located 541~m away from the target~\cite{MB_detec}. 
It is a 12.2~m diameter spherical Cherenkov detector, 
filled with the mineral oil, 
and lined with 1,280 8-inch PMTs on the wall to observe 
the Cherenkov radiation from the charged particles. 

The time and charge information of the Cherenkov rings from 
the charged particles are used to 
reconstruct charged particle momentum and particle type~\cite{MB_recon}. 
By assuming interaction is charged current quasielastic (CCQE) 
and the target nucleon is at rest, 
the neutrino energy is reconstructed (QE assumption)~\cite{MB_CCQEPRL}. 
It is vital to be able to reconstruct the neutrino energy for 
the neutrino oscillation physics. 

\subsection{MiniBooNE oscillation analysis results}

The signature of the $\numu\to\nue$ ($\numubar\to\nuebar$) oscillation is the single, 
isolated electron-like Cherenkov ring produced by the CCQE interaction. 

\beq
\numu&\stackrel{oscillation}{\longrightarrow}&\nue+n\to e^-+p~,\no\\
\numubar&\stackrel{oscillation}{\longrightarrow}&\nuebar+p\to e^++n~.\no
\eeq 

The cuts are designed to select such events. 
Both neutrino and anti-neutrino mode observed the event excesses. 
For the neutrino mode,
MiniBooNE observed the event excess only in the low energy region~\cite{MB_nu}.  
The observed excesses cannot be described by the $\nu$SM, 
so it may be the signal of new physics, 
such as Lorentz violation. 
On the other hand, for the antineutrino mode, 
the event excess is seen in the entire energy region~\cite{MB_antinu}. 
Again, the observed excesses cannot be described by the $\nu$SM~\footnote{
This analysis was done when only the half of the all antineutrino data set was available. 
Recently published full antineutrino mode data shows 
a somewhat different shape of the excess events~\cite{MB_final}.}. 
Since CPT violation naturally arises within Lorentz violation, 
this different patterns of excesses between neutrino mode and antineutrino mode 
is natural if it were caused by Lorentz violation. 
Therefore it is very interesting to take a look at their 
sidereal time distributions to find a possible Lorentz violation.

\subsection{MiniBooNE Lorentz violation analysis results}

\begin{figure}[tbp]
\begin{center}
\includegraphics[width=15cm]{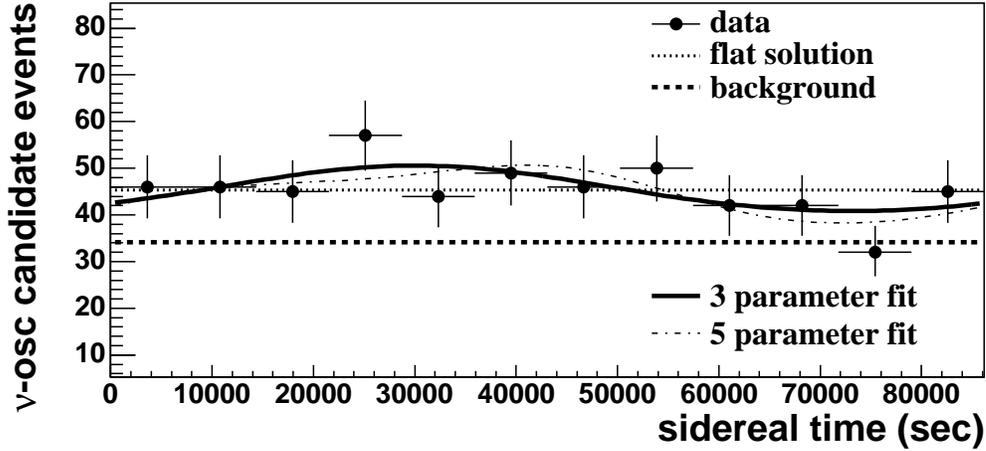}
\end{center}
\caption{\label{fig:MB_nu}
The fit results for the MiniBooNE neutrino mode low energy region. 
The plot shows the curves corresponding to the 
flat solution (dotted line),  
sidereal time dependent fits (solid and dash-dotted curves), 
together with binned data (solid marker). 
Here the fitted background is shown as a dashed line. 
Figure is taken from ~\cite{MB_LV}.
}
\end{figure}

We fit Eq.~\ref{eq:nu_prob} to MiniBooNE neutrino mode low energy $\nue$ candidate excess 
and antineutrino model excess events.  
Figure~\ref{fig:MB_nu} shows the neutrino mode low energy region fit result.  
As you see, data is quite consistent with a flat hypothesis. 
We constructed a fake data set without signal (=flat hypothesis) 
to evaluate the compatibility with a flat solution 
over the fit result by the $\De\ch^2$ method. 
It turns out data is compatible with a flat solution over a $\Dcsqnu$\%, 
and it concludes $\nue$ candidate data are consistent with no sidereal time dependence.  

\begin{figure}[tbp]
\begin{center}
\includegraphics[width=15cm]{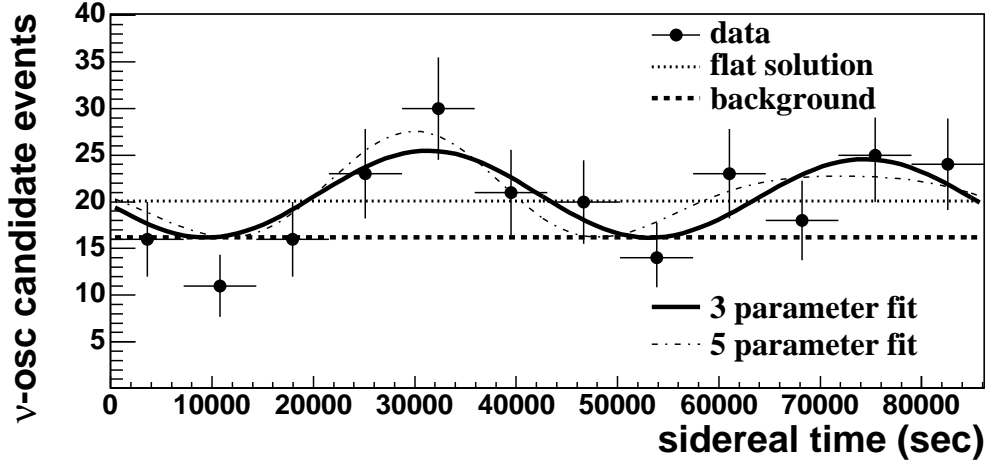}
\end{center}
\caption{
\label{fig:MB_antinu}
The fit results for the MiniBooNE antineutrino mode data. 
Notations are the same as previous figure. 
Figure is taken from ~\cite{MB_LV}.
}
\end{figure}

Figure~\ref{fig:MB_antinu} shows the antineutrino mode fit result. 
The fit result is more interesting here 
because the fit favors a sidereal time dependent solution. 
We again constructed a fake data set to find the significance of this solution, 
and it turns out that the compatibility with 
a flat solution is now only $\Dcsqnubar$\%. 
Although this is interesting, 
the significance is not high enough to claim the discovery.  

Since we didn't find the Lorentz violation, 
we set limits to fit parameters 
(=sidereal time dependent and independent amplitudes)~\cite{MB_LV}. 
From these limits one can extrapolate the limits to the SME coefficients~\cite{lv_review}. 
It turns out these limits indeed exclude 
the SME coefficients needed to explain the LSND signal. 
Therefore, there is no simple scenario to explain the LSND signal by Lorentz violation.
 
\section{Lorentz violation analysis on Double Chooz experiment}

From the MiniBooNE data analysis, 
we set limits on the $e-\mu$ oscillation channel SME coefficients. 
MINOS near detector analysis~\cite{MINOS_ND} also sets 
severe limits to some of these coefficients. 
Meantime, MINOS far detector data~\cite{MINOS_FD} and IceCube data~\cite{IceCube_LV} 
set tight limits on the $\mu-\ta$ oscillation channel SME coefficients. 
The last untested channel is the $e-\ta$ sector, 
and this can be tested by the reactor $\nuebar$ disappearance data, 
because nonzero Lorentz violating neutrino oscillation in 
the $e-\ta$ channel would contribute to the reactor neutrino disappearance. 
We analyzed data from the Double Chooz reactor experiment, 
where $\sim$4~MeV reactor $\nuebar$ are detected by 
the detector located at $\sim$1050~m away.

\begin{figure}[tbp]
\begin{center}
\includegraphics[width=12cm]{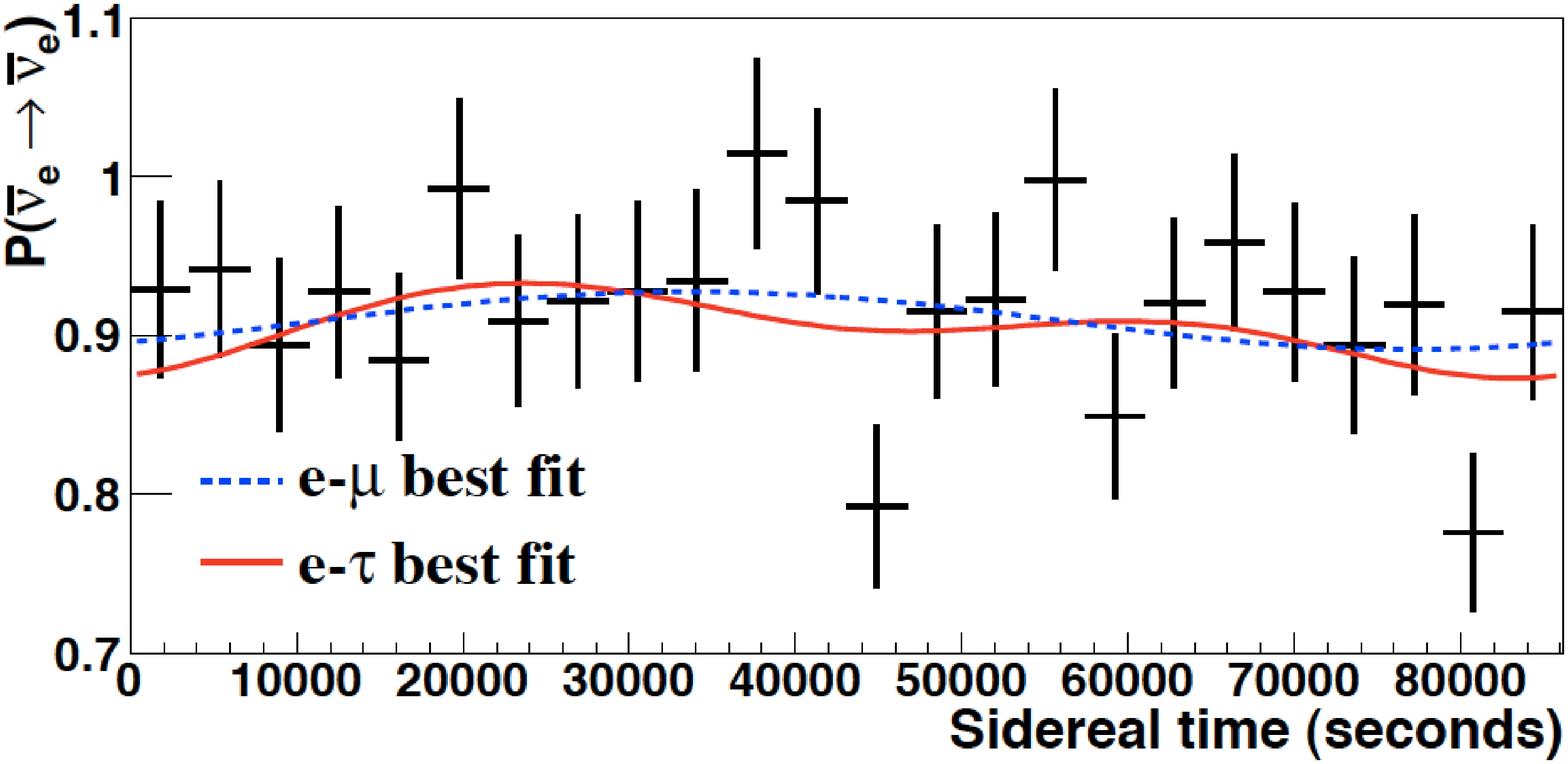}
\end{center}
\caption{\label{fig:DC}
The fit results for the reactor antineutrino data at Double Chooz far detector. 
The raio of data to simulation is overlaid with the best fit curves of 
models with Lorentz violation. 
Figure is taken from ~\cite{DC_LV}.
}
\end{figure}

Figure~\ref{fig:DC} shows the result. 
Since the reactor power varies with a day-night cycle, 
the neutrino flux is a function of the solar time (period $\tgmt$~sec) and 
it can mimic the sidereal time variation effect (period $\tsid$~sec), 
unless data taking is continuous in all one year. 
This is not the case for Double Chooz. 
However, the reactor cycle effect is simulated and taken into account in our analysis. 
 We found that the data over simulation is flat and 
the data is consistent with no sidereal time dependence. 
Therefore we set limits on $e-\ta$ sector SME coefficients. 

With the addition of this work, 
most of SME coefficients of all neutrino oscillation channels are constraint. 
Since neutrino oscillation is an interference experiment, 
as opposed to time of flight (TOF) which is a kinematic measurement, 
neutrino oscillation experiment is far more sensitive to 
small phenomena such as Lorentz violation. 
Therefore it is difficult to explain 
superluminal neutrinos observed by the OPERA experiment~\cite{OPERA_LV} 
while keeping all null Lorentz violation signals in neutrino oscillation experiments. 
Therefore, it will be challenging to detect Lorentz violation in 
the neutrino sector in any terrestrial experiments. 
In the future, astrophysical neutrinos~\cite{IceCube} may improve sensitivity 
to Lorentz violation by many orders of magnitude comapared to these limits.

\section{Conclusions}

Lorentz and CPT violation has been shown to occur in Planck-scale physics. 
There is a world wide effort to test Lorentz violation with 
various state-of-art technologies, 
including neutrino oscillations. 
LSND and MiniBooNE data suggest Lorentz violation is 
an interesting solution to neutrino oscillations. 
MiniBooNE neutrino data prefer a sidereal time independent solution, 
and MiniBooNE antineutrino data prefer a sidereal time dependent solution, 
although statistical significance is not high. Limits from MiniBooNE exclude 
simple Lorentz violation motivated scenario for LSND. 
Finally, MiniBooNE, LSND, MINOS, IceCube, and Double Chooz set sringent limits on 
Lorentz violation in neutrino sector in terrestrial level.

\section*{Acknowledgements}

I thank Jennifer Dickson for a careful reading of this manuscript. 
I also thank Jorge D\'{i}az for valuable comments.
Finally I thank to the ICHEP 2012 organizers and IUPAP C11 committee 
for the invitation to the ICHEP2012 conference.

\end{document}